\documentclass[10pt,twocolumn,twoside] {IEEEtran}

\usepackage{amsmath}
\usepackage{amssymb}
\usepackage{graphicx}
\usepackage{multirow}
\usepackage{amsfonts}
\usepackage{dsfont}
\usepackage{mathrsfs,amssymb}
\usepackage[english]{babel}
\usepackage[T1]{fontenc}
\usepackage{amsthm}
\usepackage{bbding}
\usepackage{epsfig}

\usepackage{color}

\usepackage{algorithm}
\usepackage{algorithmic}

\usepackage[font=normalsize]{caption}

\begin{document}
\begingroup

\title{Bipartite Graph based Construction of Compressed Sensing Matrices }
%
%
%


\author{Weizhi Lu, Kidiyo Kpalma and Joseph Ronsin \thanks{ The authors are with
INSA de Rennes, IETR, UMR  CNRS 6164, F-35708, France. Email:
\{weizhi.lu, kidiyo.kpalma and joseph.ronsin\}@insa-rennes.fr}}

\maketitle




\begin{abstract}


This paper proposes an efficient method to construct the bipartite graph with as many edges as possible while without introducing the shortest cycles of length equal to 4. The binary matrix associated with the bipartite graph described above presents comparable and even better phase transitions than Gaussian random matrices.
\end{abstract}


\IEEEpeerreviewmaketitle

\section{Introduction}
Compressed sensing has recently been intensively studied as a novel technique that aims to  undersample  sparse signals with low complexity \cite{Candes05}. The nonadaptive undersmapling process can be simply formulated as a system of linear equations $y=Ax$, where $x\in R^n$ is a $k$-sparse signal with at most $k$ nonzero elements, $A \in R^{m\times n}$ is a sensing matrix with $m\ll n$, and $y\in R^{m}$ is the linear observation. Given $y$ and $A$, the sparse signal $x$ can be recovered by solving the following $\ell_1$-based minimization problem
\begin{equation}
\min ||\hat{x}||_1 ~~ s.~t.~~  y=A\hat{x}
\end{equation}
if   the sensing matrix $A$ can satisfy  restricted isometry property (RIP) or null space property (NSP) \cite{Tillmann12}. The solution to formula (1) can be perfectly derived with  linear programming (LP) \cite{Boyd04}. Furthermore, for simpler computation,  a number of greedy solution algorithms are successively proposed with guaranteed performance \cite{Pati93} \cite{Daiwei09}.  Currently the explicit construction of sensing matrices well satisfying RIP or NSP is still a challenging task, since two conditions above are hard to be calculated or evaluated in practice. It is known that the random matrices with elements drawn from some general distributions, such as standard Gaussian distribution and Bernoulli distribution, can provide guaranteed sensing performance \cite{Candes05} \cite{Candes06near}. But they are computationally difficult to implement in practice. Then a few deterministic sensing matrices  are developed based on coding theory \cite{Amini11} \cite{DeVore07} \cite {Calderbank10} \cite{Liuxj13}. Specifically, the codes constructed over finite fields are collected as the columns of sensing matrices because they tend to suffer relatively small mutual correlations due to  relatively large distance/difference between each other.  However, this method is not perfect. In practice, there is no  explicit way to collect an ensemble of codes with as large average distances as possible. In addition, the codes based on primitive polynomials usually cannot constitute   arbitrary sizes of sensing matrices.

It is known that the sensing performance is inversely proportional to the maximum correlation $\mu$ between columns \cite{Donoho11}, i.e. the sparsity $k$ is upper bounded with
\begin{equation}
k<1/2(\mu^{-1}+1).
\end{equation}
Hence it is desirable if one could minimize the maximum correlations. This paper aims to approach this goal by constructing a family of deterministic matrices which is required to  have at most \emph{one} common nonzero position between two arbitrary columns. Clearly the matrices defined above  take correlation values between \emph{normalized} columns inversely proportional to the column degrees, namely the number of nonzero elements in each column. In this sense, the major task of this paper is to propose an explicit method to maximize the column degrees of the matrices defined above.

 By exploring the connection between bipartite graph and binary matrix, the construction of the desired matrices can be transformed into the construction of bipartite graphs with as many edges as possible but without cycles of length 4.  As a combinatorial optimization problem, in practice the desired bipartite graph is hard to be constructed. A greedy algorithm  initially proposed for LDPC codes and termed progressive edge-growth (PEG) \cite{Huxiaoyu05}, seems suitable for the construction of the desired bipartite graph, since it aims to  build a bipartite graph edge-by-edge without introducing short cycles at each step.  However, this method  is  imperfect on both performance and  complexity. Specifically, it has to evaluate all possible distributions of matrix column degrees for building the underlying best bipartite graph. Apparently it is a NP-hard problem.  Further, the bipartite graph constructed with PEG cannot  guarantee to have no  cycles of length 4. Thus one has to calculate the cycles of each generated bipartite graph to exclude the case of cycles of length 4.

To overcome these problems, this paper proposes a novel bipartite graph-based construction (BGC) method specially designed to construct the desired sensing matrices described above. As  will be detailed latter, the proposed BGC method outperforms PEG algorithm on both performance and complexity.  In practice, the matrices constructed with BGC method  present comparable and even better phase transition curves over Gaussian random matrices.


The rest of this paper is organized as follows. In the next Section, after introducing the  basic connection between binary matrix and bipartite graph, we describe and analyze the proposed BGC method. In Section \uppercase\expandafter{\romannumeral3}, the matrices constructed with BGC are evaluated in terms of  column degrees and phase transitions. Finally,  the paper is concluded in Section \uppercase\expandafter{\romannumeral4}.

\section{Main Results}
In this section, we first associate binary matrix with bipartite graph, and then illustrate and analyze the proposed BGC method.
\subsection{Preliminaries on bipartite graph}
  \begin{figure}[ftp]
\centering
\begin{tabular}{cc}
\includegraphics[height=60mm,width=0.19\textwidth]{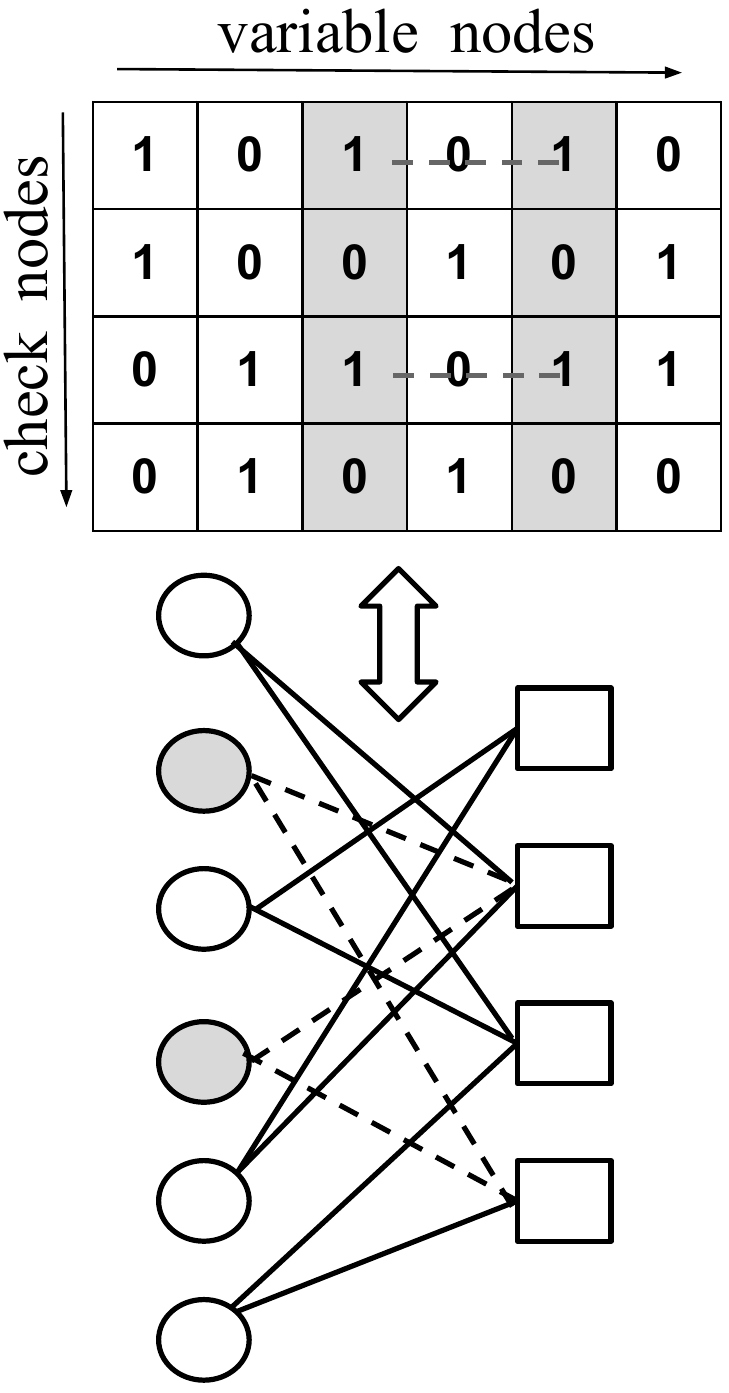}&\includegraphics[height=60mm,width=0.21\textwidth]{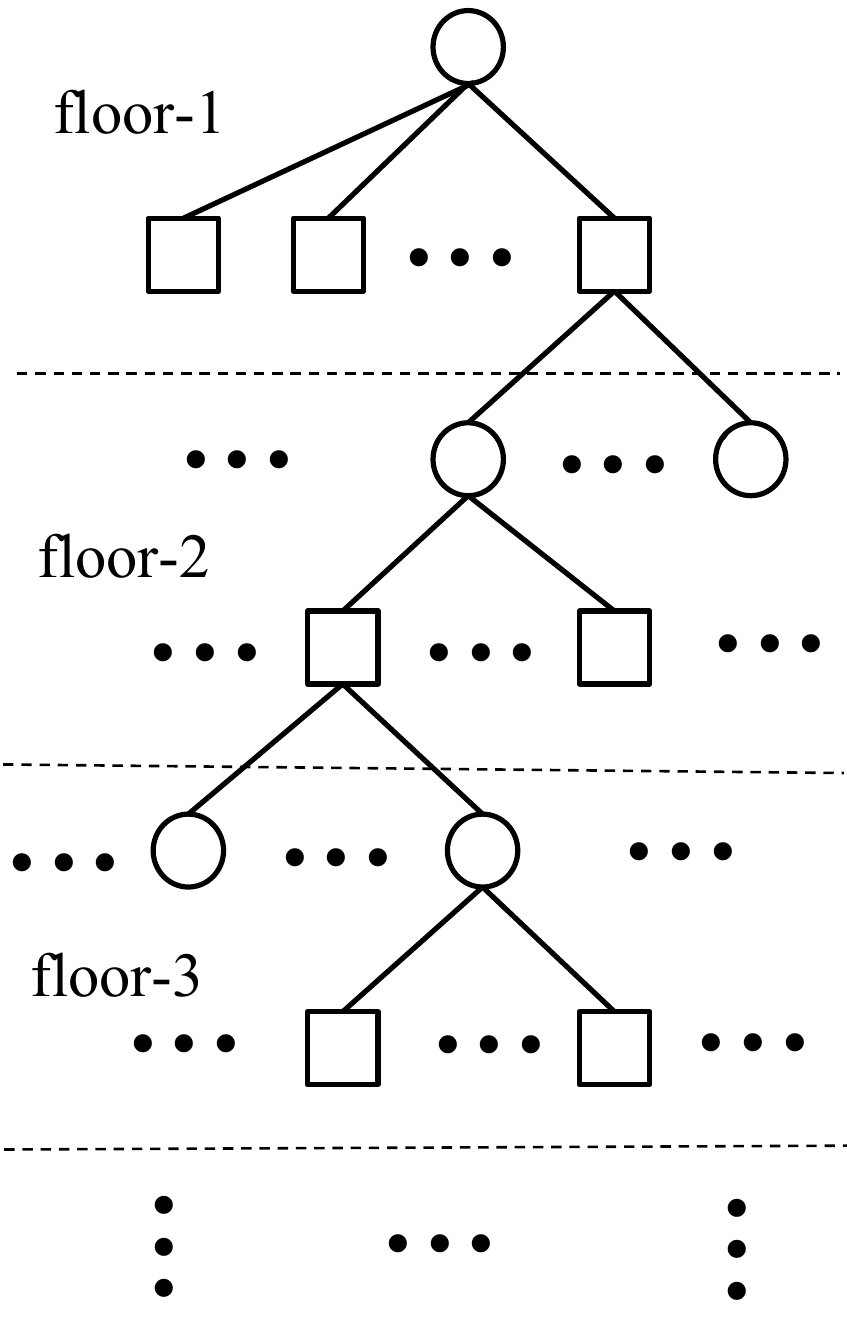}\\
(a)&(b)
\end{tabular}
\caption{ (a) An example of binary matrix being associated with bipartite
graph. The bipartite graph has variable nodes and check nodes
denoted by circles and squares, respectively. (b)
The form of subgraph expanded from a variable node. }
\end{figure}
 The bipartite graph consists of two disjoint sets of vertices and a few edges having both endpoints  from above two sets. Two sets of vertices  are typically called variable nodes and check nodes.   A  \emph{path} is defined by a sequence of vertices with two arbitrary adjacent vertices connected with an edge. A closed path starting and ending at the same vertex is called a \emph{cycle}. The length of a cycle is measured with the number of edges included in the cycle. Obviously the cycle length of a bipartite graph should be an even number not less than 4. The  length of the shortest cycle is called the \emph{girth} of bipartite graph.

 As the example shown in Figure 1(a), a sensing matrix can be  associated with a  bipartite graph  by relating columns and rows  to variable nodes and check nodes, respectively. The edges of bipartite graph are determined by the nonzero positions of sensing matrix. Hence the desired bipartite graph should have as many edges as possible. Note that  the bipartite graph will hold  cycles of length 4 through two variable nodes, if  the associated sensing matrix share two same nonzero positions between some columns. So to derive the desired sensing matrices with at most one same nonzero positions between columns, the desired bipartite graph also has to render a girth larger than 4.

 From each variable  node, as shown in Figure 1(b), a subgraph possibly with multiple floors can be generated by recursively traversing all reachable nodes through edges. Interestingly, for a given subgraph, if a check node outside the subgraph is added and connected to the root variable node, no cycles will be generated. Otherwise, a cycle of length 4 will occur if  a check node in  floor-2 is selected; likewise, a cycle of length $2i$ will appear if  a check node in  floor-$i$ is introduced.   It is thus  inevitable that the bipartite graph will have more and more shorter cycles as the edge number increases. Hence the desired bipartite graph with  as many edges as possible should have all cycles of length 6.

\subsection{Bipartite graph based construction (BGC)  method}

 As stated before, the desired bipartite graph should have all cycles of length 6. The proposed BGC method is thus developed to generate such kind of graphs with as many edges as possible. Similarly to PEG algorithm, BGC method proceeds in the way of expanding an underlying subgraph edge-by-edge. But their principles of selecting each edge are different. In each step of edge selection, BGC method prefers introducing a cycle of length 6 rather than no cycles while PEG algorithm attempts to avoid introducing cycles or short cycles. More importantly, unlike PEG, BGC does not need to enumerate all possible distributions of column degrees to seek the underlying best sensing matrix.

 The BGC method can be implemented with multiple iterations. In each iteration,  each variable node is allowed to connect at most one check node to its current  subgraph. Specifically, if the subgraph can reach floor-3 with a set of check nodes, then a check node of this floor is randomly selected to generate a cycle of length 6; otherwise,  if  current subgraph does not include all check nodes,  a check node outside of current subgraph is randomly chosen to avoid the cycle of length 4. The  procedure above is repeated  until no variable nodes have check nodes to update. Then the desired bipartite graph with as many edges as possible is derived with the following two special features:
  \begin{itemize}
  \item all existing cycles have length of 6, and  any further added edge would lead to  shorter cycles of length 4;
   \item any variable node can spread to be a subgraph with  two and only two floors  containing all check nodes.
  \end{itemize}
Two features above implies that the BGC method indeed proposes the maximum column degrees that it can greedily approach. For clarity, the proposed BGC method is sketched in  Algorithm 1.

Note that to obtain relatively uniform  column degrees, the variable nodes are updated in a random sequence in each iteration. Likewise, the candidate check nodes are also randomly selected for relatively uniform row degrees. In fact, as in \cite{Huxiaoyu05}, the row degrees can be further evened out by randomly selecting the candidate check nodes with minimum number of adjacent variable nodes. However, it is not applied in the proposed method, since in practice  it  does not lead to noticeable advances on maximizing the number of edges.

\begin{algorithm}[htp]
\caption{ BGC  method}
\label{alg:Framwork}
\textbf{Initializations}:  Let $\mathcal{C}$ and $\mathcal{V}$  refer to the set of $m$ check nodes and  the set of $n$ variable nodes, respectively. $\mathcal{I} $ is defined as the set of variable nodes still to be updated in current bipartite graph, which is initialized as $\mathcal{I}=\mathcal{V}$. $\mathcal{I}_i$  indicates the $i$-th element of $\mathcal{I} $. Searching for new edges will stop if $\mathcal{I} $ turns to be empty.  Three subsets $\mathcal{C}_i$ with subscript $1\leq i\leq 3$ are further defined to contain the check nodes appearing on the $i$-th floor of  subgraph. If the subgraph has no check nodes on  floor-$i$, the corresponding $\mathcal{C}_i=\varnothing$. The selected edges are collected in the set $\mathcal{E}$ which is initialized as $\varnothing$.
\begin{algorithmic}[1]
\FOR {$i$=1 to $m$}
\IF{$\mathcal{I}=\varnothing $ }
\STATE break; // terminate the program and output the edge set $\mathcal{E}$;
\ENDIF
\FOR{$j$=1 to $|\mathcal{I}|$}
\STATE Try to expand a subgraph from variable node $\mathcal{I}_j$ to  floor-3 with current edge set $\mathcal{E}$; and the the check nodes on the $k$-th floor are collected in the empty-initialized set $\mathcal{C}_k$, $1\leq k\leq 3$;
\IF{$\mathcal{C}_3=\varnothing$ }
\IF{$\mathcal{C}\backslash\{\mathcal{C}_1\bigcup\mathcal{C}_2\}\neq \varnothing$}
\STATE Introduce a new edge $(\mathcal{I}_j, c)$ to the edge set $\mathcal{E}$ by $\mathcal{E}=\mathcal{E}\bigcup (\mathcal{I}_j, c)$, where $c$ is a check node randomly selected from the set $\mathcal{C}\backslash\{\mathcal{C}_1\bigcup\mathcal{C}_2\}$;
\ELSE
\STATE Exclude the variable node  $\mathcal{I}_j$ from  $\mathcal{I} $, namely $\mathcal{I}=\mathcal{I}\backslash\mathcal{I}_j$;
\ENDIF
\ENDIF
\IF{$\mathcal{C}_3\neq\varnothing$}
\STATE Introduce a new edge $(\mathcal{I}_j, c)$ to the edge set $\mathcal{E}$ by $\mathcal{E}=\mathcal{E}\bigcup (\mathcal{I}_j, c)$, where $c$ is a check node randomly selected from the set $\mathcal{C}_3$;
\ENDIF
\ENDFOR
\ENDFOR
\end{algorithmic}

\end{algorithm}
 At the end of this subsection, three crucial problems concerning the practical implementation of BGC method are discussed as follows.

\subsubsection{Performance and complexity} the proposed BGC method outperforms PEG algorithm on both performance and complexity. Precisely, in performance, BGC method can provide a distribution of larger average column degrees as shown in the following Table \uppercase\expandafter{\romannumeral1}.  As for complexity, BGC shows two obvious advantages. First,  it only needs to be given the size of target matrix, while besides matrix size, PEG can proceed only with a given distribution of column degrees. In other words, theoretically PEG has to enumerate all possible distributions of column degrees to find the underlying best matrix. Second,  PEG algorithm cannot avoid  the case of girth equal to 4, and so additional computation has to be introduced to evaluate the girth of each matrix constructed with PEG; in contrast, this case does not happen in BGC method. Secondly, during the spreading of each subgraph, BGC needs to spread  at most 3 floors, while PEG has to expand  as far as possible.
\vspace{3pt}
\subsubsection{Bounds for column degrees} \emph{i) upper bound}: the maximum edge number of bipartite graph with girth larger than 4 has been early studied as a combinatorial problem  in \cite{Naor2005} and related references therein. However, empirically these asymptotic results  are far away from the real  values that BGC or PEG can achieve. Here we provide a more accurate method to tackle this problem.  Considering an ideal case where the generated matrix is regular, i.e. both column degrees and row degrees are uniform, its uniform column degree $d$ can be approximately estimated with
    \begin{equation}
    \mathop{\text{argmin}}\limits_d ~\{|d+d(dn/m-1)(d-1)-m|\},
    \end{equation}
     since we can  regard $
    d+d(dn/m-1)(d-1)= m
    $ from the fact that each subgraph finally spread two floors in which all check nodes are included. Note that theoretically it cannot be ensured that the regular  matrix described above really exists. For instance, the solution $d$ to formula (3) is usually not an integer provided a pair of $(m,n)$.    But compared to  pervious impractical  theoretical estimation \cite{Naor2005}, it is  much closer to the real values as shown in Table \uppercase\expandafter{\romannumeral1}. \emph{ ii) lower bound}: it is necessary to point out that the matrices constructed with BGC probably have some columns of degrees equal to 1, when the column size $n$ is very small or the compression rates $n/m$ is considerably large. Obviously for overcomplete sensing matrices, it is necessary to render most column  degrees larger than 1  to avoid producing same columns. In practice, as  shown in the simulations, this condition can be well satisfied as $n$ or $m/n$ increases. \vspace{3pt}
\subsubsection {Binary and Ternary matrices} with given bipartite graph, in practice we can construct two types of matrices: deterministic binary matrix and random ternary matrix by taking the nonzero values as  1 and  random binary values $\pm 1$, respectively. Both of them share the same distribution on the magnitudes of column correlation. But the same performance cannot be assured in practice. Recall that, empirically, the sensing performance is  sensitive to the sign of sparse signals \cite{Donoho10112009}.  It is thus reasonable to conjecture that  the \emph{unsigned} binary matrix and the \emph{signed} ternary matrix also probably differ in performance.  This conjecture is confirmed with the following simulations.


\section{Simulations}

\begin{table*}[htp]
\caption{For sensing matrices with diverse sizes, the \emph{average} column degrees  provided by BGC are compared with the maximum\emph{ uniform} column degrees achieved by PEG. The theoretical estimation derived with formula (3), here denoted as 'Th', is illustrated as well.}
\centering
\renewcommand\arraystretch{1.0}
\begin{tabular}{|c|c|c|c|c|c|c|c|c|c|c|c|c|}

\hline
\multicolumn{3}{|c|}{m/n}&0.1 &0.2&0.3 &0.4&0.5 &0.6& 0.7&0.8 &0.9&1\\
\hline\hline

 \multirow{9}{*}{n}&\multirow{3}{*}{\rotatebox{90}{100}}& BGC& \textbf{1.45}& \textbf{ 2.37}& \textbf{3.2}&\textbf{3.99}&\textbf{4.68}& \textbf{5.27}& \textbf{5.93}& \textbf{6.55}& \textbf{7.21}&\textbf{7.69}\\
&& PEG&  1&   2&3&  3& 4& 5& 5&6& 7& 7\\
&&Th&1.44&2.00&2.50&2.96&3.39&3.79&4.18&4.56&4.90&5.25\\

\cline{2-13}
&\multirow{3}{*}{\rotatebox{90}{500}}& BGC& \textbf{2.68}&  \textbf{4.45}& \textbf{5.99}&\textbf{7.40}& \textbf{8.67}& \textbf{9.86}& \textbf{10.99}& \textbf{12.08}& \textbf{13.07}&\textbf{14.21}\\
&& PEG&  2&   4&5&  7& 8& 9& 10& 11& 12& 13\\
&&Th&2.12&3.12&3.98&4.76&5.48&6.16&6.80&7.41&8.00&8.57\\

\cline{2-13}

 & \multirow{3}{*}{\rotatebox{90}{1000}}& BGC& \textbf{3.48}&  \textbf{5.82}&7\textbf{.80} &\textbf{9.59}&\textbf{11.26} &\textbf{12.78}& \textbf{14.36}& \textbf{15.75}&\textbf{17.06} &\textbf{18.45}\\
&& PEG&  3&   5&7 & 9& 11 &12& 14 &15& 16& 17\\
&&Th&2.55&3.83&4.91&5.88&6.78&7.63&8.43&9.20&9.93 &10.64\\

\hline

\end{tabular}
\end{table*}

\begin{figure}[t]
\centering
\begin{tabular}{c}
\includegraphics[width=0.42\textwidth]{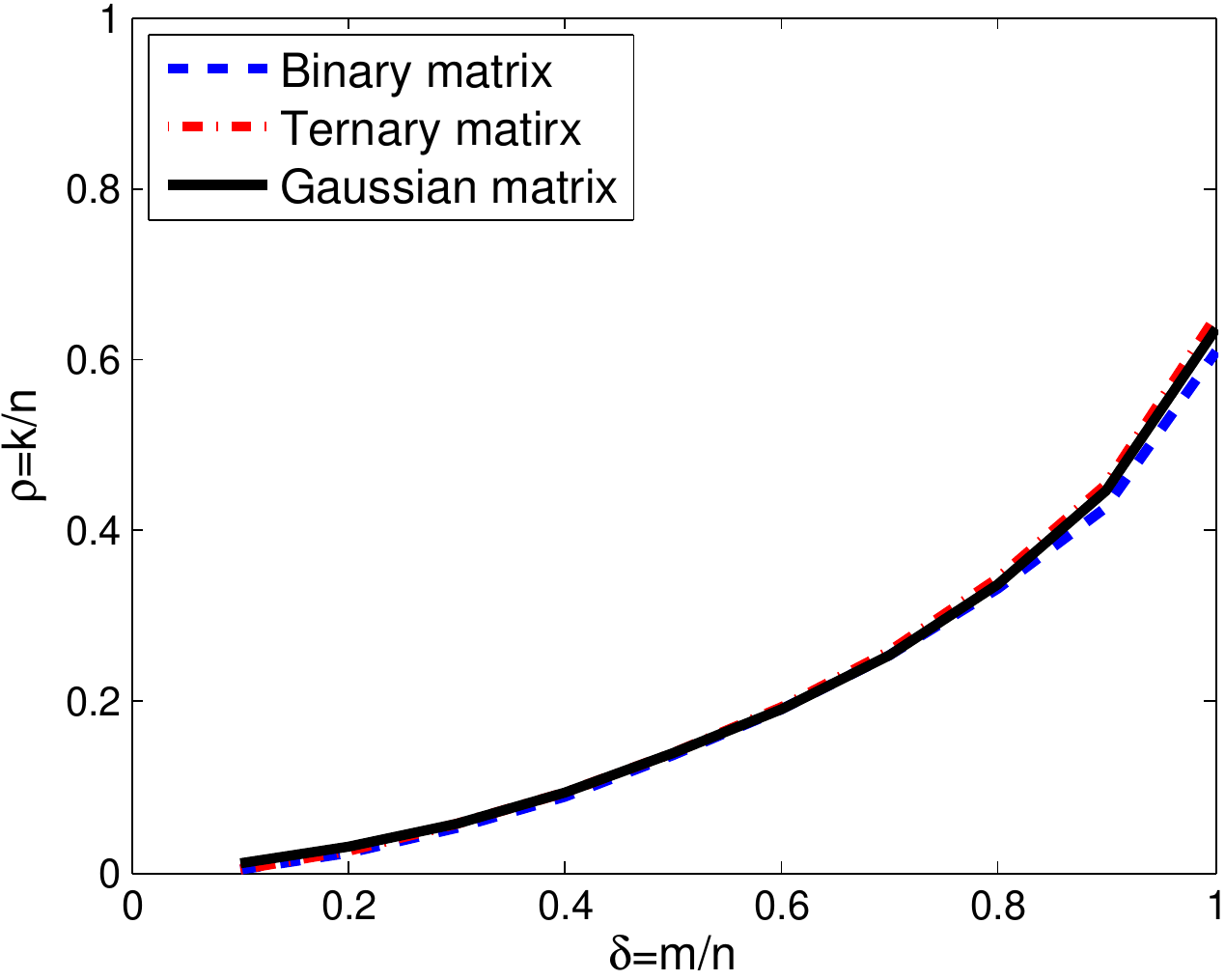}\\ (a) \\ \includegraphics[width=0.42\textwidth]{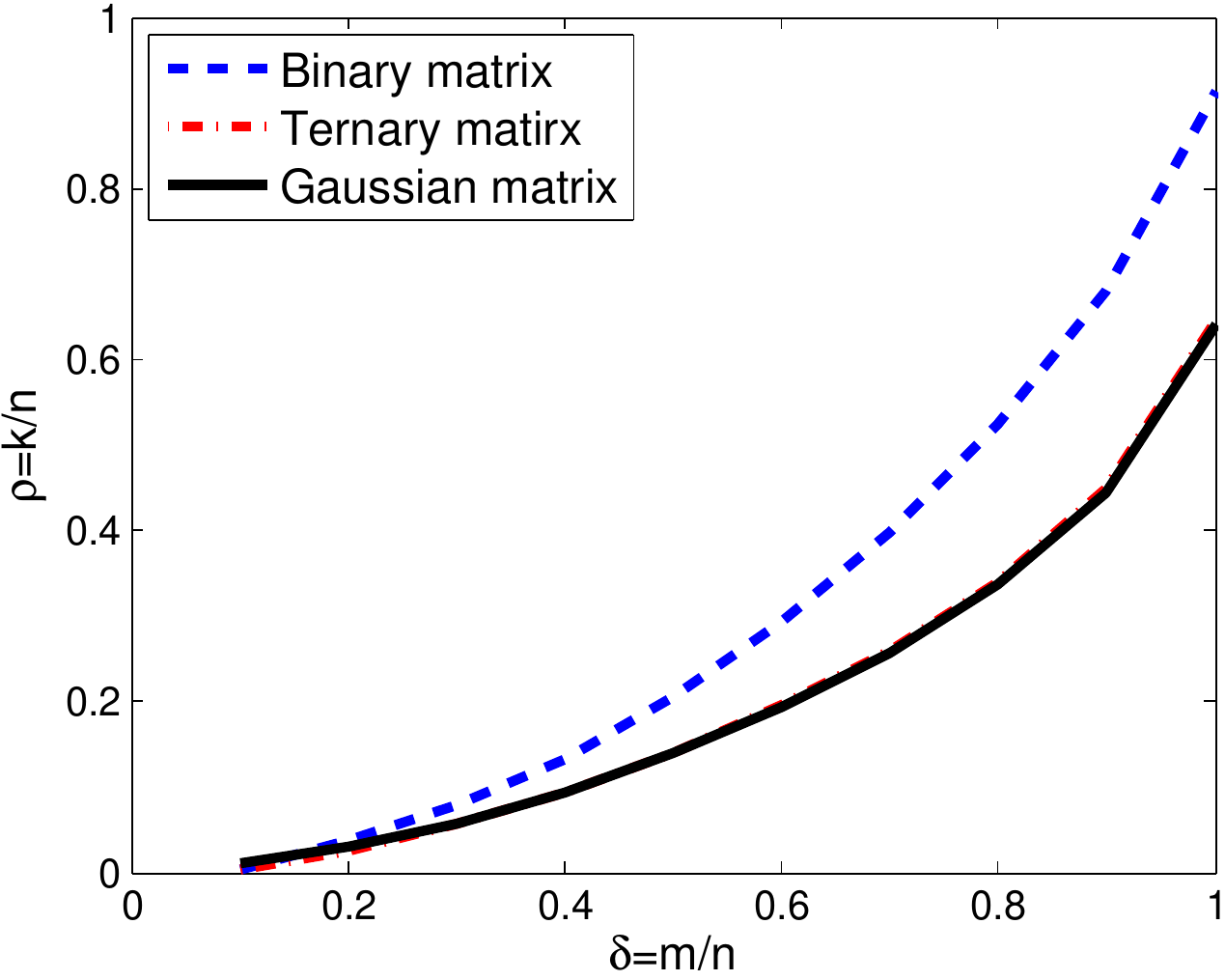}\\ (b)

\end{tabular}
\caption{The phase transitions of binary matrices and ternary matrices constructed by BGC, as well as Gaussian random matrix. The sparse signals have nonzero elements  being $\pm1$ equiprobably in (a), and being $ 1$ in (b).}
\end{figure}
This section first compares the maximum column degrees achieved by  BGC and  PEG, and then evaluates  the phase transitions  of  matrices constructed with BGC.
\subsection{Maximum column degrees}

The maximum column degrees achieved by BGC and PEG are compared in Table \uppercase\expandafter{\romannumeral1}. Simultaneously, the theoretical estimation derived with formula (3) is also illustrated.  Apparently BGC obtains larger average column degrees  over PEG. The requirement of avoiding  column degrees equal to 1 is also well satisfied as $n $ or $m/n$ increases. Note that PEG is only used to seek the desired matrices of uniform column degrees, because  we cannot enumerate all possible distributions of column degrees.  In practice PEG tends to provide larger average column degrees if the column degrees are distributed non-uniformly. For instance, for the case of $n=500$ and $m/n=0.1$ in Table \uppercase\expandafter{\romannumeral1}, PEG in fact can achieve an average column degree 2.5 by having  half of the columns with degree 2, and  the rest columns with degree 3. Even so,  PEG can hardly  achieve the performance level of BGC, since we do not know the better  distribution of column degrees  derived with BGC.  Moreover, it can be observed that the practical matrices constructed with PEG or BGC  outperform the theoretical estimation determined with  formula (3). This estimation error can be partially explained by the fact that, the theoretical estimation is derived under the ideal assumption that both row degrees and column degrees are uniform, while the condition above is not  followed by PEG or BGC. In particular,   BGC has no requirement for uniform column or row degrees, and PEG here is exploited to merely produce uniform column degrees.


\subsection{Phase transitions}

The phase transitions \cite{Donoho10} are illustrated in Figure 2. Besides the binary matrices and ternary matrices constructed with BGC, Gaussian random matrix are evaluated as well. These matrices are of column size $n=300$. The curves are depicted by the maximum nonzero number $k $ allowed by  the correct LP decoding rates larger than $99\%$.  The decoding rates are measured with $||\hat{x}-x||_2/||x||_2$.  Note that  $\delta=m/n$ ranges discretely  from 0.1 to 1 in nine equal steps. Each point is derived from 1000 simulations.  In each simulation, the ternary matrix has nonzero elements randomly taking $\pm1$ signs. Similarly to \cite{Donoho10112009}, here we test both \emph{signed} and \emph{unsigned} sparse signals with nonzero elements i.i.d drawn from the sets  $\{ \pm1\}$ and $\{ 1\}$, respectively.

Due to the low resolution of Figure 2, it is necessary to point out that the as $\delta$ increases, the performance order is Ternary matrix>Gaussian matrix>Binary matrix in Figure 2(a), and Binary matrix>Ternary matrix>Gaussian matrix in Figure 2(b). In terms of the average performance, obviously the deterministic binary matrix is more attractive in practice.

\section{Conclusion}

This paper has proposed a bipartite graph based construction method, called BGC method to deterministically construct binary or ternary sensing matrices with  column correlations as small as possible. This method is developed by equivalently regarding the desired matrix as a bipartite graph with as many edges as possible but without short cycles of length 4. The matrices constructed with BGC show comparable and even better performance over Gaussian random matrices. Note that, in practice the PEG algorithm initially proposed for the construction of LDPC codes can also be applied to construct such kind of graphs. But this method can hardly seek the underlying best distribution of column degrees unless we could enumerate all possible distributions. The proposed BGC method not only successfully works out this problem but significantly reduces computation complexity. In the future, it can be easily adapted to generate more hardware-friendly  matrices of quasi-cyclic structure.

\bibliographystyle{IEEEtran}
\bibliography{egbib}

\begin{thebibliography}{10}
\providecommand{\url}[1]{#1}
\csname url@samestyle\endcsname
\providecommand{\newblock}{\relax}
\providecommand{\bibinfo}[2]{#2}
\providecommand{\BIBentrySTDinterwordspacing}{\spaceskip=0pt\relax}
\providecommand{\BIBentryALTinterwordstretchfactor}{4}
\providecommand{\BIBentryALTinterwordspacing}{\spaceskip=\fontdimen2\font plus
\BIBentryALTinterwordstretchfactor\fontdimen3\font minus
  \fontdimen4\font\relax}
\providecommand{\BIBforeignlanguage}[2]{{%
\expandafter\ifx\csname l@#1\endcsname\relax
\typeout{** WARNING: IEEEtran.bst: No hyphenation pattern has been}%
\typeout{** loaded for the language `#1'. Using the pattern for}%
\typeout{** the default language instead.}%
\else
\language=\csname l@#1\endcsname
\fi
#2}}
\providecommand{\BIBdecl}{\relax}
\BIBdecl

\bibitem{Candes05}
E.~Candes and T.~Tao, ``Decoding by linear programming,'' \emph{IEEE
  Transactions on Information Theory}, vol.~51, no.~12, pp. 4203 -- 4215, dec.
  2005.

\bibitem{Tillmann12}
A.~M. {Tillmann} and M.~E. {Pfetsch}, ``{The Computational Complexity of the
  Restricted Isometry Property, the Nullspace Property, and Related Concepts in
  Compressed Sensing},'' \emph{ArXiv e-prints}, May 2012.

\bibitem{Boyd04}
S.~Boyd and L.~Vandenberghe, \emph{Convex Optimization}.\hskip 1em plus 0.5em
  minus 0.4em\relax Cambrige university press, March 2004.

\bibitem{Pati93}
Y.~Pati, R.~Rezaiifar, and P.~Krishnaprasad, ``Orthogonal matching pursuit:
  recursive function approximation with applications to wavelet
  decomposition,'' in \emph{Conference Record of The Twenty-Seventh Asilomar
  Conference on Signals, Systems and Computers}, nov 1993, pp. 40 --44 vol.1.

\bibitem{Daiwei09}
W.~Dai and O.~Milenkovic, ``Subspace pursuit for compressive sensing signal
  reconstruction,'' \emph{IEEE Transactions on Information Theory}, vol.~55,
  no.~5, pp. 2230--2249, 2009.

\bibitem{Candes06near}
E.~Candes and T.~Tao, ``Near-optimal signal recovery from random projections:
  Universal encoding strategies?'' \emph{IEEE Transactions on Information
  Theory}, vol.~52, no.~12, pp. 5406 --5425, Dec. 2006.

\bibitem{Amini11}
A.~Amini and F.~Marvasti, ``Deterministic construction of binary, bipolar, and
  ternary compressed sensing matrices,'' \emph{IEEE Transactions on Information
  Theory}, vol.~57, no.~4, pp. 2360 --2370, april 2011.

\bibitem{DeVore07}
R.~A. DeVore, ``Deterministic constructions of compressed sensing matrices,''
  \emph{Journal of Complexity}, vol.~23, no. 4-6, pp. 918 -- 925, 2007.

\bibitem{Calderbank10}
R.~Calderbank, S.~Howard, and S.~Jafarpour, ``Construction of a large class of
  deterministic sensing matrices that satisfy a statistical isometry
  property,'' \emph{IEEE Journal of Selected Topics in Signal Processing},
  vol.~4, no.~2, pp. 358--374, 2010.

\bibitem{Liuxj13}
X.-J. Liu and S.-T. Xia, ``Constructions of quasi-cyclic measurement matrices
  based on array codes,'' in \emph{IEEE International Symposium on Information
  Theory Proceedings (ISIT)}, 2013, pp. 479--483.

\bibitem{Donoho11}
D.~Donoho and X.~Huo, ``Uncertainty principles and ideal atomic
  decomposition,'' \emph{IEEE Transactions on Information Theory}, vol.~47,
  no.~7, pp. 2845--2862, Nov 2011.

\bibitem{Huxiaoyu05}
X.-Y. Hu, E.~Eleftheriou, and D.~Arnold, ``Regular and irregular progressive
  edge-growth {Tanner} graphs,'' \emph{IEEE Transactions on Information
  Theory}, vol.~51, no.~1, pp. 386 --398, jan. 2005.

\bibitem{Naor2005}
A.~Naor and J.~Verstra\"{e}te, ``A note on bipartite graphs without
  2k-cycles,'' \emph{Comb. Probab. Comput.}, vol.~14, no. 5-6, pp. 845--849,
  Nov. 2005.

\bibitem{Donoho10112009}
D.~L. Donoho, A.~Maleki, and A.~Montanari, ``Message-passing algorithms for
  compressed sensing,'' \emph{Proceedings of the National Academy of Sciences},
  vol. 106, no.~45, pp. 18\,914--18\,919, 2009.

\bibitem{Donoho10}
D.~Donoho and J.~Tanner, ``Precise undersampling theorems,'' \emph{Proceedings
  of the IEEE}, vol.~98, no.~6, pp. 913--924, 2010.

\end{thebibliography}

\end{document}